\documentclass[12pt]{article}
\usepackage{wrapfig}
\usepackage{epsfig}
\usepackage{hyperref}
\usepackage{amssymb}
\usepackage{amsmath,mathtools}
\usepackage{amsfonts}
\usepackage{graphicx}
\usepackage{latexsym}
\usepackage{slashed}
\usepackage{multirow}
\usepackage{fixmath}
\usepackage{color}
\usepackage{stackrel}
\usepackage{txfonts}
\usepackage{centernot}
\usepackage{mathabx}
\usepackage{bbm}
\usepackage[low-sup]{subdepth}

\newcommand{\bea}{\begin{eqnarray}}
\newcommand{\eea}{\end{eqnarray}}
\newcommand{\bite}{\begin{itemize}}
\newcommand{\eite}{\end{itemize}}

\newcommand{\appropto}{\mathrel{\vcenter{
  \offinterlineskip\halign{\hfil$##$\cr
    \propto\cr\noalign{\kern2pt}\sim\cr\noalign{\kern-2pt}}}}}

\date{}

\begin{document}
\title{
\vspace*{-1.25cm}
{\bf Absence of strong CP violation}}

\author{G.~Schierholz\\[1em] 
Deutsches Elektronen-Synchrotron DESY,\\ Notkestr. 85, 22607 Hamburg, Germany\\[1em] Email: gerrit.schierholz@desy.de}

\maketitle
%\vspace*{-0.5cm}

\begin{abstract}
Quantum Chromodynamics admits a CP violating contribution to the action, the $\theta$ term, which is expected to give rise to a nonvanishing electric dipole moment of the neutron. Despite intensive search, no CP violations have been found in the strong interaction. This puzzle is referred to as the strong CP problem. There is evidence that CP is conserved in the confining theory, to the extent that color charges are totally screened for $\theta > 0$ at large distances. It is not immediately obvious that this implies a vanishing dipole moment though. With this Letter I will close the gap. It is shown that in the infinite volume hadron correlation functions decouple from the topological charge, expressed in terms of the zero modes. The reason is that hadrons have a limited range of interaction, while the density of zero modes vanishes with the inverse root of the volume, thus reducing the probability of finding a zero mode in the vicinity to zero. This implies that CP is conserved in the strong interaction.
\end{abstract}

\section{Introduction}

We consider Quantum Chromodynamics (QCD) in Euclidean space-time. Lattice regularization in finite volume $V$ is assumed, which is the standard technique for the nonperturbative definition of the theory, although continuum notation is applied. The consequences for infinite volume will be discussed later. The most general action for two mass-degenerate $u$ and $d$ quarks is given by  
\begin{equation}
  S = \int d^4x\, \Big[\frac{1}{2g^2}\, \textrm{Tr}\, F_{\mu\nu} F_{\mu\nu} - i\, \theta_1\, \epsilon_{\mu\nu\rho\sigma} \frac{1}{32\pi^2}  \textrm{Tr}\, F_{\mu\nu} F_{\rho\sigma} + \sum_{q=u,d}\bar{q}\left(-i\slashed{D}+\mathcal{M} \right)q \Big] \,,
  \label{action1}
\end{equation}
with
\begin{equation}
  \mathcal{M}=\frac{1}{2}(1-\gamma_5)\, M + \frac{1}{2}(1+\gamma_5)\, M^\dagger \,, \quad M=m\, e^{\,i\,\theta_2/2}\,.
  \label{action2}
\end{equation}
The normalization is chosen so that $\theta_1$ is an angular variable,
\begin{equation}
  Q = \epsilon_{\mu\nu\rho\sigma} \,\frac{1}{32\pi^2}  \int d^4x  \,\textrm{Tr}\, F_{\mu\nu} F_{\rho\sigma} \in \mathbb{Z} \,,
  \label{Q}
\end{equation}
where $Q$ is the topological charge. The second term on the right-hand side of (\ref{action1}), the so-called $\theta$ term, and the mass term (\ref{action2}) violate CP for nonzero phases $\theta_1$ and $\theta_2$. The quark fields $q$ can always be redefined by a global flavor singlet axial rotation,
\begin{equation}
  q \rightarrow e^{-i\,(\alpha/4) \gamma_5}\, q \,, \quad \bar{q} \rightarrow \bar{q}\, e^{-i\,(\alpha/4) \gamma_5} \,,
\end{equation}
which changes the phases by $\theta_1 \rightarrow \theta_1 - \alpha$ and $\theta_2 \rightarrow \theta_2 + \alpha$ due to a change of the path integral measure, leaving the sum $\theta_1 + \theta_2$ unchanged. Thus, a change of $\theta_1$ is eqivalent to a change of the phase of the quark mass matrix $\theta_2$ by the negative amount. The noninvariance of the path integral measure is due to the anomaly. In the following we consider the two extreme cases, $\theta_2 = 0$ and $\theta_1 = 0$, and denote the complementary phases, $\theta_1$ and $\theta_2$, synonymously by $\theta$.

A primary CP violating observable is the electric dipole moment of the neutron, $d_n$, which is a measure of the separation of positive and negative electric charges within the neutron. So far no electric dipole moment has been found. The current upper bound is $|d_n| < 1.8 \times 10^{-13} e\,\textrm{fm}$~\cite{Abel:2020pzs}, which is commonly taken as an indication that $\theta$ is anomalously small. An estimate of $\theta$ requires to know $d_n/\theta$ though. As the theory does not have an enhanced symmetry when $\theta$ is sent to zero, it would be natural for $\theta$ to take any nonzero value between $-\pi$ and $+\pi$. This puzzle is generally referred to as the strong CP problem. It should be noted that actually only the imaginary part of the phase factor $\exp\{i\,\theta Q\}$ and its fermionic counterpart in the path integral violate CP, while the real part results in the free energy, which is the key function in the study of phases and phase transitions.

The strong CP problem is considered one of the biggest unsolved problems in the elementary particles field. It is deeply rooted in the topological properties of the theory. A nonvanishing value of $\theta$ is expected to have a significant impact on the QCD vacuum~\cite{Callan:1979bg,tHooft:1981bkw}. In fact, there may be infinitely many `$\theta$ vacua', all separated by phase boundaries~\cite{Cardy:1981qy}. This raises the question why Nature chose $\theta=0$. In the first place, this suggests to look for correlations between $\theta$ and other parameters of the theory that are anthropically constrained. A `solution' was found by examining the renormalization group flow of the running coupling $\alpha_s$ depending on $\theta$ in the Yang-Mills theory~\cite{Nakamura:2019ind,Nakamura:2021meh}. It turned out that $\alpha_s$ and $\theta$ are driven to a fixed point, $1/\alpha_s = \theta = 0$, in the infrared exhibiting linear confinement. This led us to conclude that confinement implies CP invariance of the strong interaction~\cite{Nakamura:2019ind}. %This does not contradict~\cite{Shifman:1979if}, as confinement is lost for nonvanishing values of $\theta$. Actually, strong CP invariance only demands that hadron observables are an even function of $\theta$.

Then why this Letter? There is a loophole in this argument. Confinement is a property of the QCD vacuum, which is characterized by nonvanishing vacuum condensates, such as the gluon condensate~\cite{Schierholz:2025tns}, all of which are CP-even. CP violation, on the other hand, arises from correlations of the $\theta$ term with the electromagnetic current of the neutron, for example. This is a separate problem that explores different aspects of the vacuum. It is expected that QCD is analytic in $\theta$~\cite{Vafa:1984xg} at $\theta = 0$, which results in a smooth phase transition or crossover. This will carry over to hadron observables. It is therefore possible to find a stable neutron and a nonvanishing dipole moment for $|\theta| > 0$. Current lattice calculations of the electric dipole moment also show this. The range of $\theta$ may be unexpectedly large. If $\theta$ is increased the color charges are gradually screened. The Debye screening length was found to be inversely proportional to $|\theta|$~\cite{Schierholz:2022wuc}. While the color charge will be screened for any $|\theta| > 0$ at large distances, the nucleon will disintegrate into quarks and gluons once the screening length has become smaller than the nucleon radius, which might be the case for $|\theta| \gtrsim 0.2$ only~\cite{Nakamura:2021meh}. Such a phenomenon has actually been observed shortly above the finite temperature phase transition, where the screening length is inversely proportional to $(T-T_c)$~\cite{Matsui:1986dk}. Although this picture is very appealing, further research is needed to verify it. In this Letter I will concentrate on the CP violating part of the path integral and show that the dipole moment, and the like, vanish for any (small) value of $\theta$ that keeps the quarks bound together. The mechanism here is that the $\theta$ term decouples from hadron correlation functions in the large volume limit. %This should dispel any doubts that CP is conserved in the strong interaction. The Letter may be regarded an independent proof of the matter.

\section{Topology, axial anomaly and index theorem}

The path integral splits into quantum mechanically disconnected sectors of topological charge $Q$. On the lattice this happens at lattice spacings $a \lesssim 0.05\,\textrm{fm}$. The topological charge is obtained by differentiating the action (\ref{action1}) with respect to $\theta$ (either $\theta_1$ or $\theta_2$). This has two solutions, the field-theoretic expression (\ref{Q}) and the so-called fermionic definition of $Q$, arising from the anomaly. Expressing the fermion fields in terms of eigenfunctions of the Dirac operator and integrating out the Grassmann variables, the latter reads~\cite{Fujikawa:1980eg}\\
\begin{equation}
  Q = -\sum_{i} \int d^4x \; u^\dagger_i(x) \gamma_5 u_i(x)\,, 
  \label{top}
\end{equation}
where \{$u_i(x)$\} are the normalized zero-mode eigenvectors of the Dirac operator,
\begin{equation}
   \slashed{D}\, u_i(x) = 0\,, \quad \int d^4x\, u^\dagger_i(x) u_i(x) = 1 \,,
\end{equation}
of which $n_+$ eigenvectors have positive chirality, $\gamma_5\, u_i(x)=u_i(x)$, and $n_-$ have negative chirality, $\gamma_5\, u_i(x)=-u_i(x)$. Thus $Q = n_- - n_+$, which is known as the Atiyah-Singer index theorem. On the lattice, the definition (\ref{top}) of the topological charge is realized by~\cite{Hasenfratz:1998ri} 
\begin{equation}
  Q = \frac{1}{2} \, \textrm{Tr}\,\gamma_5 D_N = - \frac{1}{2} \, \textrm{Tr}\,\gamma_5 (2-D_N) = - \frac{1}{2} \sum_\lambda (2-\lambda)\, \left(u_\lambda, \gamma_5 u_\lambda\right) \,, 
  \label{topover}
\end{equation}
where $u_\lambda$ are the eigenvectors of the overlap Dirac operator, $D_N\, u_\lambda = \lambda\, u_\lambda$~\cite{Neuberger:1997fp}. The eigenvalues of $D_N$ lie on the circle $\lambda=1-e^{i \phi}$. There are two types. Real eigenvalues, $\lambda = 0$ (the zero modes) and $\lambda = 2$ with $\gamma_5 u_\lambda = \pm u_\lambda$, as well as complex eigenvalues with $\gamma_5 u_\lambda = u_{\lambda^*}$ and $\left(u_\lambda, \gamma_5 u_\lambda\right) = 0$. It follows that
\begin{equation}
Q = - \sum_{\{\lambda = 0\}} (u_0,\gamma_5 u_0)\,,
\label{topf}
\end{equation}
in agreement with the formal expression (5). We emphasize that the definition (\ref{topf}) is determined exclusively by the zero modes. We shall use this definition of $Q$ throughout this Letter. It should not depend on the particular boundary taken in the large volume limit. Physically, the zero-mode eigenfunctions are found to be strongly correlated with instantons of the appropriate charge~\cite{DeGrand:2000gq}. 

%It is readily seen that both definitions of $Q$, (\ref{top}) and (\ref{topover}), are identical. 

The essential point now is that any gauge field configuration of positive (negative) topological charge $Q$ has exactly $n_-$ ($n_+$) zero modes, but never zero modes of both chiralities, that is to say $|n_+ -n_-|=n_+ + n_-$. In mathematical language: $\textrm{dim ker}\, i \slashed{D} = |n_+ - n_-|$. This so-called `vanishing theorem' was proven in two dimensions~\cite{Ansourian:1977qe,Nielsen:1977aw}, and for self-dual configurations in four dimensions~\cite{Brown:1977bj,Carlitz:1978xu}. A proof for generic metrics and (`gauge') connections in two and four dimensions was given in~\cite{Maier}.
%Mathematically, references~\cite{Ansourian:1977qe,Nielsen:1977aw,Maier} put the `vanishing theorem' on the same level as the Atiyah-Singer index theorem.
This is a surprising result that takes some getting used to. A few familiar results may help. In QCD the `vanishing theorem' applies to the effective theory. By equating the $\theta$ term and the mass term $\bar{q}\mathcal{M}q$ in the action~(\ref{action1}), following~\cite{Leutwyler:1992yt}, we get
\begin{equation}
  Z = \sum_Q e^{i\, \theta\, Q}\, Z_Q \equiv e^{-V\, F(\theta)} \,,
  \label{zf}
\end{equation}
where $F(\theta)$ is the free energy, which depends on the combination $m e^{i \theta/2}$ only (for two quark flavors). Treating the mass term as a perturbation, this leads to 
\begin{equation}
  F = - \frac{1}{V} \log Z = - 2\, \Sigma \,\textrm{Re} \left(m e^{i\theta/2}\right) + O(m^2)\,, \quad \Sigma = - \langle \bar{q} q\rangle \,.
  \label{free}
\end{equation}
The partition function $Z_Q$ is obtained by Fourier expanding the right-hand side of (\ref{zf}), which gives
\begin{equation}
  Z_Q= I_{|Q|}^2(V \Sigma m) - I_{|Q|+1}(V \Sigma m)\,I_{|Q|-1}(V \Sigma m) \,.
  \label{ZQ}
\end{equation}
In the chiral limit this reduces to
\begin{equation}
  Z_Q= \left(\frac{V \Sigma m}{2}\right)^{2 |Q|} \frac{1}{|Q|!\,(|Q|+1)!} \,,
\end{equation}
which is consistent with the fermion determinant and implies the occurrence of exactly $|Q|$ topological zero modes. This result is an integral part of chiral perturbation theory and its spectral manifestations~\cite{Leutwyler:1992yt}. Zero modes of mixed chirality, $n_+ + n_- > |n_+ - n_-|$, would be severely suppressed for small quark masses by an additional factor of $m^{2(n_+ + n_- - |n_+ - n_-|)}$ in the fermion determinant, besides being of measure zero. For $Q=0$, for example, the factor would be $m^{2(n_+ + n_-)}$. The `vanishing theorem' has been confirmed in numerous lattice simulations in two~\cite{Chiu:1998bh} and four dimensions at zero~\cite{Ilgenfritz:2007xu,Chiu:2011dz,DiGiacomo:2015eva} and finite temperature~\cite{Chen:2022fid,Chiu}, up to spatial volumes of $(5\, \textrm{fm})^3$. A proof through `reduction' is provided by the gradient flow~\cite{Luscher:2010iy}, which is an infinitesimal, differentiable and invertible RG-type transformation that connects lattice gauge fields to semi-classical, self-dual multi-instanton solutions at larger flow times $t$, $\langle E\rangle_{t\rightarrow \infty}V \simeq |Q|\, 8\pi^2$~\cite{Nakamura:2021meh}. It is known that $\partial Q/\partial t \equiv \partial (n_- - n_+)/\partial t = 0$~\cite{Luscher:2010iy,Nakamura:2021meh}. Evidently, it is not possible to change a zero mode,
\begin{equation}
  \left(u_0,\gamma_5 u_0\right) = \pm 1 \,,
\end{equation}
into a continuous mode,
\begin{equation}
  \left(u_\lambda,\gamma_5 u_\lambda \right) = 0 \,, \quad |\lambda| > 0 \,,
\end{equation}
by an infinitesimal change of $t$. It follows that the number of zero modes is independent of the flow time, from the multi-instanton regime to present lattice cut-offs, which means that either $Q = n_-$ or $Q = -n_+$. %In summary, the topological charge is mapped one-to-one to the number of zero modes.

The `vanishing theorem', and the fact that the vacuum splits into disconnected sectors of topological charge $Q$, has notable consequences. To be mentioned first is the cluster property, which states that truncated correlators tend to zero faster than any power of distance. It has been argued that it is violated in the presence of instantons and anti-instantons~\cite{Luscher:1979yd}. Consider the topological charge density
\begin{equation}
  q(x) = \epsilon_{\mu\nu\rho\sigma} \,\frac{1}{32\pi^2} \,\textrm{Tr}\, F_{\mu\nu} F_{\rho\sigma} \,.
\end{equation}
The cluster decomposition theorem requires that 
\begin{equation}
  \langle q(x) \,q(y) \rangle \stackrel[|x-y| \rightarrow \infty]{}{=} \langle q(x) \rangle\, \langle q(y) \rangle = \langle q(x) \rangle^2 \,.
  \label{cluster}
\end{equation}
The argument would be correct if the vacuum was in a mixed state of instantons and anti-instantons~\cite{Bohm:2001yx}. This is, however, not the case. The `vanishing theorem' states that the vacuum accommodates exactly $|Q|$ zero modes. In effect, the cluster property is restored by restricting the theory to a Hilbert space built on two degenerate vacua. A similar situation is met in the nonlinear $O(3)$ model in two dimensions~\cite{Iwasaki:1981yq}. A further consequence concerns the dilute instanton gas approximation of the QCD vacuum. The model is characterized by a charge distribution consisting of a convolution of two independent Poisson distributions for instantons and anti-instantons, thus superimposing charges of both chiralities. This is incompatible with the `vanishing theorem'. The dilute instanton gas might serve as a toy model for the low-lying nonzero modes though.
%single instanton and anti-instanton solutions isolated from each other. The presence of both fermion chiralities excludes the model, besides having an infrared divergence. 

\section{Hadron correlators, electric dipole moment and beyond}

In the $\theta$ vacuum an $n$-point correlation function of operators $\mathcal{O}_1, \dots ,\mathcal{O}_n$ takes the form
\begin{equation}
  \langle \mathcal{O}_1 \cdots \mathcal{O}_n \rangle_\theta = \langle e^{i\,\theta\,Q}\, \mathcal{O}_1 \cdots \mathcal{O}_n \rangle \,. %= \frac{1}{Z}\, \sum_Q  e^{i\,\theta\,Q} \, P(Q) \; \langle \mathcal{O}_1 \cdots \mathcal{O}_n \rangle_Q \,,
\end{equation}
%where $P(Q)$ denotes the probability for charge $Q$ (at $\theta=0$), with $\sum_Q P(Q)=1$. The respective definition of $Z$ results from the context.
The electric dipole moment of the nucleon is given by
\begin{equation} 
  \vec{d}_n = \dfrac{\int d^3\vec{x} \: d^3\vec{y} \: e^{i \vec{p} \vec{x}}\, \langle N(\vec{x},x_0)\, \vec{y}\, J_0(\vec{y},y_0)\, \bar{N}(0) \rangle_\theta}{\int d^3x \: e^{i \vec{p} \vec{x}}\, \langle N(\vec{x},x_0) \, \bar{N}(0) \rangle_\theta}\,,
\label{dipole1}\end{equation}
where $J_0$ is the time component of the electromagnetic current and $x_0 \gg y_0 \gg 0$. The support of $\vec{y}$ is limited to the spatial size of the nucleon. Appropriate traces of the correlators $\, \langle \cdots \rangle\,$ are to be taken, which we have omitted. In practice, large ($\gg$) means about a fermi apart. Treating the $\theta$ term as a perturbation at first order, we obtain
\begin{equation} 
  \vec{d}_n = - i\,\theta \; \dfrac{\int d^3\vec{x} \: d^3\vec{y} \: e^{i \vec{p} \vec{x}}\, \langle \big(\sum_i \int d^4z\, u^\dagger_i(z)\gamma_5 u_i(z)\big)\,N(\vec{x},x_0)\, \vec{y}\, J_0(\vec{y},y_0)\, \bar{N}(0) \rangle}{\int d^3x \: e^{i \vec{p} \vec{x}}\, \langle N(\vec{x},x_0) \, \bar{N}(0) \rangle}\,,
  \label{dipole2}
\end{equation}
where the (global) topological charge $Q$ in the numerator, being an integer, is given by the integral over the zero-mode eigenvectors (\ref{top}). The nonzero modes contribute only indirectly through the fermion determinant.  

\begin{figure}[t!]
  \vspace*{-0.25cm}
  \begin{center}
    \includegraphics[width=9cm]{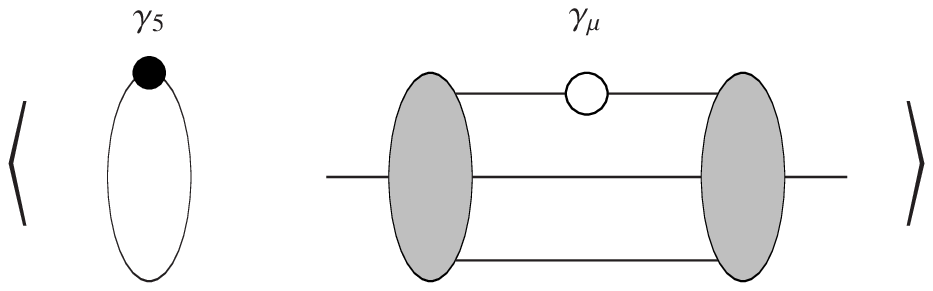}
  \end{center}
  \vspace*{-0.75cm}
  \caption{The disconnected diagram of the zero-mode flavor singlet pseudoscalar density and the nucleon isovector electromagnetic current.}
  \label{fig1}
\end{figure}

Our expression (\ref{dipole2}) for the correlation function is in total agreement with the evaluation of the dipole moment based on the anomalous chiral Ward identity~\cite{Guadagnoli:2002nm}. The result is sketched in Fig.~\ref{fig1}. It is characterized by a quark-line disconnected diagram, where the disconnected part is a quark loop with $\gamma_5$ insertion. The quark propagator can be written
\begin{equation}
  S(x,y)=\sum_\lambda \frac{u_\lambda(x) u_\lambda^\dagger(y)}{i\lambda + m}
  \label{prop}
\end{equation}
for small quark masses. With $(u_\lambda,\gamma_5 u_\lambda) =0$ for $\lambda \neq 0$, only the zero modes contribute to the loop, just like in (\ref{dipole2}). 

\begin{figure}[b!]
  \vspace*{0.25cm}
  \begin{center}
\includegraphics[width=5.5cm]{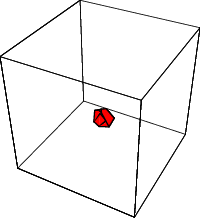} \hspace*{1cm}   \includegraphics[width=5.5cm]{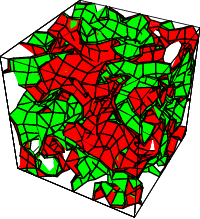}
  \end{center}
  %\vspace*{0.25cm}
  \caption{The distribution of topological charge density $q(x)$ for $Q=1$ in a given time slice on the $12^3\times 24$ lattice at lattice spacing $a=0.125 \, \textrm{fm}$~\cite{Koma:2005sw}, differentiated by positive (red) and negative (green) charge densities. Both figures show isosurfaces of $|q|=0.0005$. The left panel displays the zero mode. The right panel shows the sum over all eigenvectors. Results are from a cluster analysis~\cite{Ilgenfritz:2007xu}. The spatial extent of the eigenfunctions can be extracted from the inverse participation ratio as well. See Sec. 4.}
  \label{fig2} \vspace*{0.25cm}
\end{figure}

A characteristic of the nucleon correlation function (\ref{dipole1}) is that it extends over little more than the spatial size of the nucleon and spreads over Euclidean times of a few fermi only. A nonvanishing dipole moment arises from the interaction of the electromagnetic current with one of the zero modes (\ref{dipole2}). In a series of papers~\cite{Ilgenfritz:2007xu,Koma:2005sw,Ilgenfritz:2006gc,Ilgenfritz:2007ua,Ilgenfritz:2009qw} it has been shown that the zero-mode eigenvectors are highly localized, in contrast to the nonzero modes. This is demonstrated in Fig.~\ref{fig2} for a representative gauge field configuration of charge $Q = 1$~\cite{Koma:2005sw}, showing the (lattice) topological charge density
\begin{equation}
  q(x) = - \frac{1}{2} \sum_\lambda (2-\lambda) u^\dagger_\lambda(x) \gamma_5 u_\lambda(x) \,.
\end{equation}
The left panel shows the contribution of the zero mode in a given time slice. The right panel shows the total contribution summed over all eigenvectors, which fills the entire volume. The zero-mode density $|u_0(x)|^2$ drops by an order of magnitude at a distance of a fraction of a fermi from its center. Thus, the dipole moment (\ref{dipole2}) can be visualized by a disconnected diagram of two rather local operators, as shown in Fig.~\ref{fig1}. The result will depend on the probability of finding a zero mode within the interacting range. If the zero modes are out of range, it does not matter whether they have positive or negative chirality (and the associated topological charges are positive or negative), due to the absence of long-range interactions. In this case the nucleon is exposed only to the nonzero modes that make up the sea, and correlators composed of quark propagators restricted to these modes reproduce ordinary hadron correlators at small quark masses. Incidentally, the nonzero modes can be converted to pairs of positive and negative chirality, $\sum_{\lambda \neq 0} u_\lambda^\dagger u_\lambda = (1/2)\sum_{\lambda \neq 0} (u_{+ \,\lambda}^\dagger u_{+ \,\lambda} + u_{- \,\lambda}^\dagger u_{- \,\lambda})$ with $u_{\pm \,\lambda} = (1/2)\, (u_{\lambda} \pm u_{\lambda^*})$, which `neutralize' each other and leave no $\theta$ dependence behind in the fermion determinant~\cite{Leutwyler:1992yt}. 

\begin{figure}[b!]
  \vspace*{-0.9cm}
  \begin{center}
    \includegraphics[width=13cm]{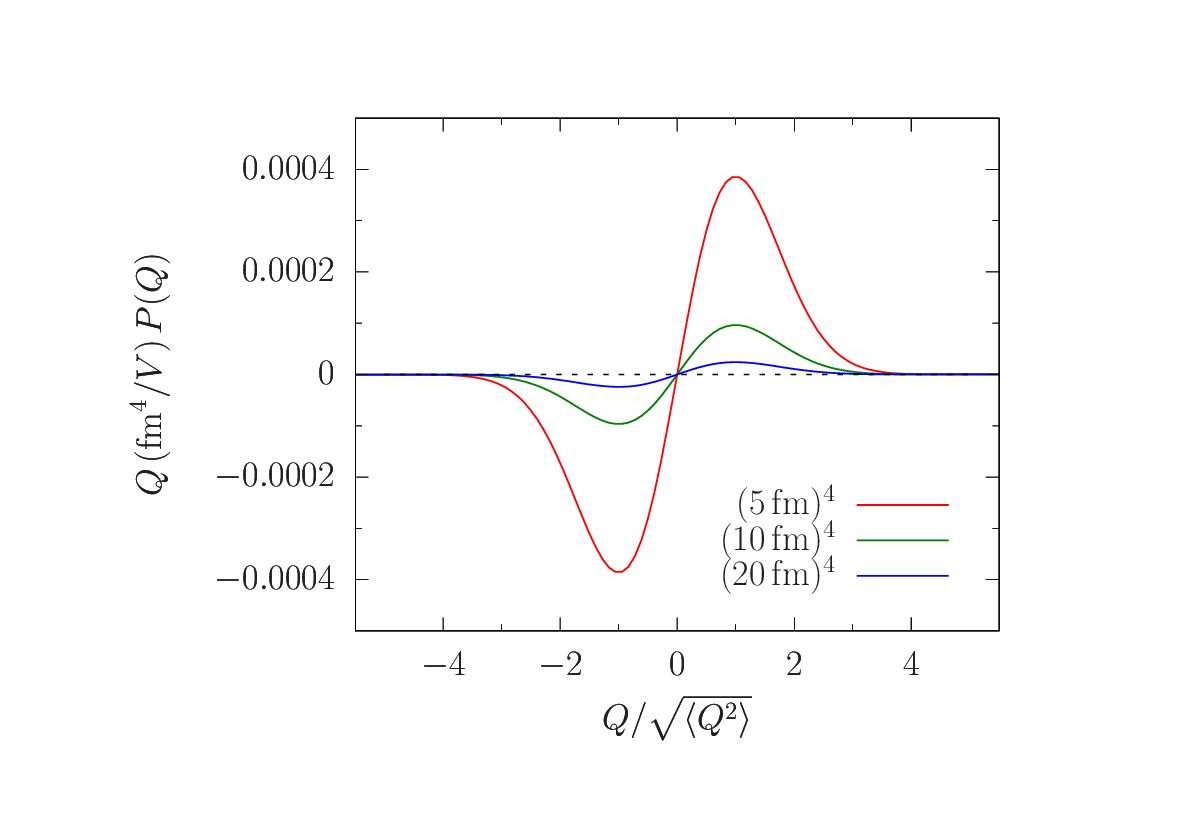}
  \end{center}
  \vspace*{-1.25cm}
  \caption{The fraction of zero modes to be expected in a subvolume of one $\textrm{fm}^4$ as a function of $Q = \pm n$ for total volumes $V$ of $(5 \,\textrm{fm})^4$, $(10 \,\textrm{fm})^4$ and $(20 \,\textrm{fm})^4$. Scale setting is described in the text.}
  \label{fig3}
\end{figure}

In the following we call the number of zero modes $n$. That can be either $n=n_+$ or $n=n_-$, according to the `vanishing theorem'. We can assume that the topological susceptibility
\begin{equation}
  \chi_t = \frac{\langle Q^2 \rangle}{V} \equiv \frac{\langle n^2 \rangle}{V}
  \label{ts}
\end{equation}
is independent of the space-time volume $V$. The decisive factor for the magnitude of $d_n$ is the density of zero modes. We assume that the probability function $P(Q)$ for charge $Q$ can be described by a Gaussian distribution,
\begin{equation}
  P(Q) \equiv \frac{Z_Q}{Z} =\frac{1}{\sqrt{2\pi \langle Q^2\rangle}} \;e^{-Q^2/2\langle Q^2\rangle}\,, \quad |Q|=n \,.
  \label{gc}
\end{equation}
Minor corrections should not matter. The density of zero modes is then given by
\begin{equation}
  \frac{\langle n \rangle}{V} = \sqrt{\dfrac{2}{\pi}} \, \dfrac{\sqrt{\langle Q^2\rangle}}{V} = \sqrt{\dfrac{2}{\pi}} \,\sqrt{\dfrac{\chi_t}{V}} \,,
  \label{density}
\end{equation}
which vanishes with the inverse square root of the volume.

\begin{figure}[b!]
  \vspace*{-0.75cm}
  \begin{center}
    \includegraphics[width=12.5cm]{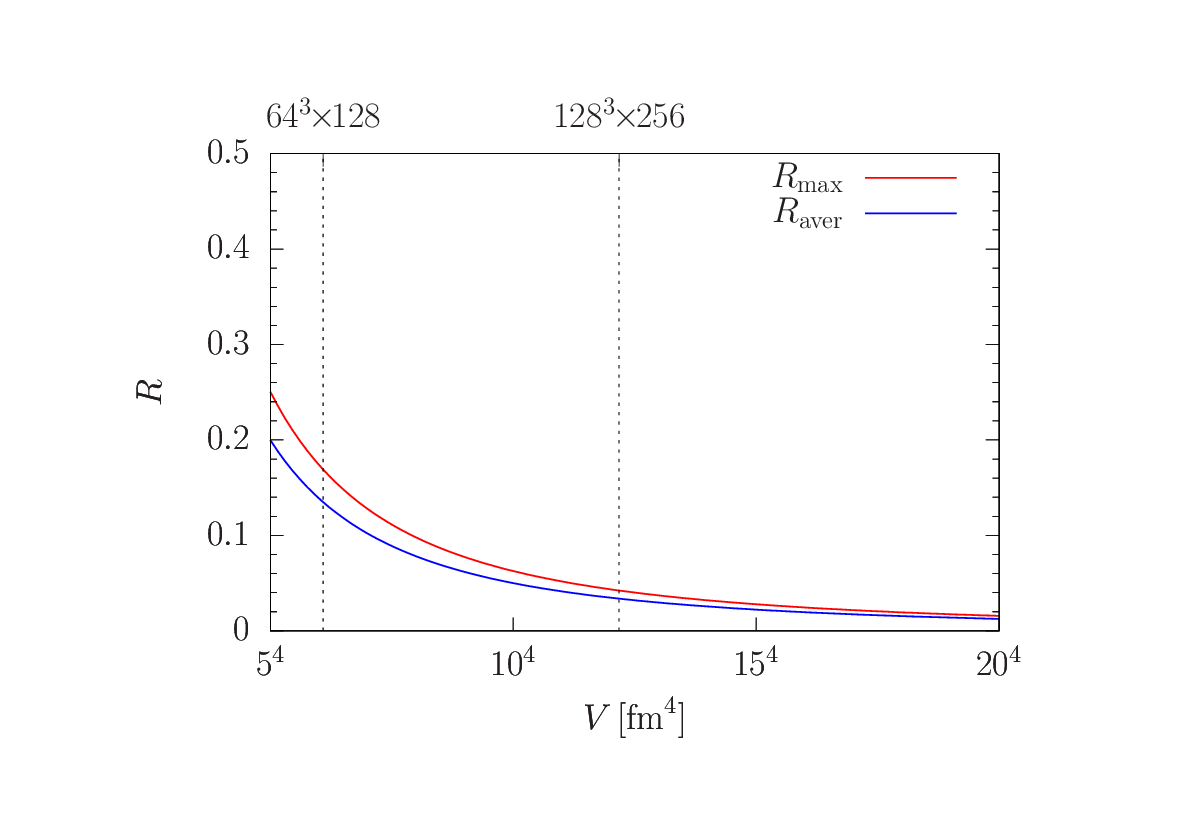}
  \end{center}
  \vspace*{-1.0cm}
  \caption{The probability for the maximum and average amount of zero modes in the subvolume $V_0$ as a function of the total volume $V$. Corresponding volumes of a $64^3\times 128$ and a $128^3\times 256$ lattice at lattice spacing $a=0.08\,\textrm{fm}$ are shown at the top.}
  \label{fig4}
\end{figure}

To make any quantitative statement about the density of zero modes (\ref{density}), we need to know $\chi_t$ in physical units. Chiral perturbation theory predicts $\chi_t=(79\,\textrm{MeV})^4$ for $N_f=2+1$ quark flavors at the physical point using the quenched (Yang-Mills) result  $\chi_t^{YM} = (f_\pi^2/6)\,(m_{\eta^\prime}^2 + m_{\eta}^2 -2 m_K^2) = (179\,\textrm{MeV})^4$ as input. The fraction of zero modes to be expected on average in a subvolume of one $\textrm{fm}^4$ is given by $|Q|\, (\textrm{fm}^4/V) \, P(Q)$ ($|Q|=n$). The result is shown in Fig.~\ref{fig3} for three lattice volumes, $V = (5 \,\textrm{fm})^4$, $(10 \,\textrm{fm})^4$ and $(20 \,\textrm{fm})^4$. By increasing the lattice volume from $(5 \,\textrm{fm})^4$ to $(10 \,\textrm{fm})^4$ and $(20 \,\textrm{fm})^4$ the fraction of zero modes shrinks by a factor of $4$ and $16$, respectively. We can assume that the strong interactions are largely confined to a box of about $(2.5 \,\textrm{fm})^4$, the `femto universe'. We take that as our reference volume $V_0$. To put the size of $V_0$ into perspective, in a recent calculation of the electric dipole moment on the $64^3\times 128$ lattice at lattice spacing $a=0.08 \, \textrm{fm}$~\cite{Alexandrou:2020mds} that would correspond to a $32^4$ sublattice. The exact size of $V_0$ does not matter for our final conclusion. The all-important question then is, what is the probability of finding a zero mode in the reference volume $V_0$? According to Fig.~\ref{fig3} the maximum of $n$ is at about $|Q| \approx \sqrt{\langle Q^2\rangle}$. Thus, on average, we can expect at most $R_{\textrm{max}}=\sqrt{\langle Q^2 \rangle}\,(V_0/V)$ zero modes within the interacting range of the electromagnetic current. For the average number (\ref{density}) the result is  $R_{\textrm{aver}} = \sqrt{2\langle Q^2 \rangle/\pi}\,(V_0/V)$. In Fig.~\ref{fig4} we show both numbers as a function of total volume $V$. The numbers are compared with a fictious lattice calculation on the $64^3\times 128$ and the $128^3\times 256$ lattice at lattice spacing $a=0.08 \, \textrm{fm}$. While on the $64^3\times 128$ lattice the probability for finding a zero mode within range is $\approx 15\%$, the probability drops to $\approx 4\%$ already on the $128^3\times 256$ lattice. Higher powers of the $\theta$ term in (\ref{dipole2}) will not change the general picture, because there is a much lower probability for finding more than one zero mode. We can exclude long-range correlations~\cite{Leutwyler:1992yt}. The final result is that the dipole moment vanishes in the infinite volume proportional to
\begin{equation}
  |d_n| \, \propto \, \sqrt{\frac{\chi_t}{V}} \, |\theta| \,.
  \label{result}
\end{equation}
Corresponding estimates apply to the CP violating pion-nucleon coupling constant $\bar{g}_{\pi NN}$~\cite{Crewther:1979pi} and the like.

Now to potential CP violating contributions beyond the dipole moment. Shifman, Vainshtein and Zacharov~\cite{Shifman:1979if} argued that the $\theta$ term implies a nonvanishing value of
\begin{equation}
  \epsilon_{\mu\nu\rho\sigma} \,\frac{1}{32\pi^2} \,\textrm{Tr}\, \langle F_{\mu\nu} F_{\rho\sigma} \rangle \equiv \frac{1}{V} \, \langle Q \rangle =  \theta \; \frac{\partial}{\partial \theta}\, \frac{1}{V} \langle Q\rangle \, \big|_{\,\theta=0} + O(\theta^3) \,,
  \label{svz}
\end{equation}
which, though not directly measurable, gives rise to CP violation in physical transition amplitudes. However, the authors overlooked the fact that (\ref{svz}) vanishes in the large volume limit, like the density (\ref{density}). Imposing the cluster property upon the vacuum, the path integral splits into twofold degenerate sectors of positiv and negative topological charge (see the discussion at the end of Sec.~2 and Ref.~\cite{Iwasaki:1981yq}). Considering either one of them, say $Q \geq 0$, we find 
\begin{equation}
  \frac{\partial}{\partial \theta}\, \frac{1}{V} \langle Q\rangle \, \big|_{\,\theta=0} = - i \, \frac{1}{V} \left(\langle Q^2\rangle - \langle Q\rangle^2\right)\, \big|_{\,\theta=0} \,.
  \label{deriv}
\end{equation}
If we now express the topological charge in terms of the zero modes and follow the same reasoning that led us to (\ref{result}), we obtain for the large volume
\begin{equation}
%  \begin{split}
%    \langle Q^2\rangle &= \;\:
%   \langle \sum_i \int d^4x\, u_i^\dagger(x)\gamma_5 u_i(x) \, \sum_j \int d^4y\, u_j^\dagger(y)\gamma_5 u_j(y)\rangle \\
%    &\!\!\!\underset{V \rightarrow \infty}{=} \langle |Q|\rangle + \langle \sum_i \int d^4x \, u_i^\dagger(x)\gamma_5 u_i(x)\rangle^2\\
%    &= \;\; \langle |Q|\rangle +\langle Q\rangle^2 \,.
%  \end{split}
  \langle \sum_i \int d^4x\, u_i^\dagger(x)\gamma_5 u_i(x) \, \sum_j \int d^4y\, u_j^\dagger(y)\gamma_5 u_j(y)\rangle \simeq \langle n\rangle + \langle \sum_i \int d^4x \, u_i^\dagger(x)\gamma_5 u_i(x)\rangle^2 \,,
  \label{deriv2}
\end{equation}
where $\langle n\rangle$ reflects the contact term. This finally results in the asymptotic expression
\begin{equation}
  \epsilon_{\mu\nu\rho\sigma} \,\frac{1}{32\pi^2} \,\textrm{Tr}\, \langle F_{\mu\nu} F_{\rho\sigma} \rangle \simeq - i \, \theta \, \sqrt{\frac{2}{\pi}} \,\sqrt{\frac{\chi_t}{V}} 
  \label{FFdual}
\end{equation}
to leading order in $\theta$.
%If we had not imposed the cluster property, but summmed over positive and negative charges, we would get $\langle Q \rangle = 0$ in (\ref{deriv}) and (\ref{deriv2}), without changing the final result.
Another criticism is that the authors of~\cite{Shifman:1979if} implicitly assume that the theory confines for nonvanishing values of $\theta$, which is not the case for general values of $\theta$~\cite{Nakamura:2021meh}. Taking full account of the dynamical properties of the vacuum, our preliminary results suggest that $(1/V)\, \langle Q\rangle = - i \, (1/V)\, \text{\large$\wp$}(\theta;0,g_3)$ on larger volumes for $\theta \gg 0$, where \text{\large$\wp$} is Weierstrass' elliptic function~\cite{mathematica} and $g_3$ is a parameter that determines its period. This is a much stronger result than (\ref{FFdual}).

%Finally, this tells us that CP is conserved in the strong interactions. To be precise, that means in the confining regime of the theory.  

\section{Summary and conclusion}

The strong CP problem is largely a question of why the neutron electric dipole moment $d_n$ is unnaturally small. To lowest order in $\theta$, the electric dipole moment arises from the nucleon matrix element of the electromagnetic current weighted with the (global) topological charge $Q$. The topological charge is conveniently expressed by the zero-mode eigenvectors. Only a zero mode can generate a CP violating, $\theta$-odd contribution to the effective electromagnetic current, in accord with the anomalous Ward identity~\cite{Guadagnoli:2002nm}. The eigenvectors are found to be highly local~\cite{Ilgenfritz:2007xu,Koma:2005sw}, and appear to be attached to instantons and anti-instantons~\cite{DeGrand:2000gq}. Thus, we are dealing with the interaction of the nucleon with a point-like source, leaving aside that the color charges will eventually be screened and the nucleon disintegrate at larger values of $\theta$~\cite{Nakamura:2021meh,Schierholz:2022wuc}. There are exactly $|Q|$ zero modes. If the topological susceptibility $\chi_t= \langle Q^2\rangle/V$ is independent of the volume, the density of zero modes vanishes with the inverse root of the volume in the thermodynamic limit. A byproduct is that the probability of finding a zero mode within a correlation length of the boundary decreases with the fourth root of the volume. As a result, the probability of finding a zero mode within the interacting range of the nucleon decreases rapidly as the physical volume is increased, leading to $d_n = 0$ in the infinite volume limit. %Similar results hold for any other CP violating coupling of the nucleon, as well as for the gluon condensate~\cite{Shifman:1979if}. 

The electric dipole moment of the neutron has been the subject of many investigations in the past, using effective  theory~\cite{Crewther:1979pi,Guo:2012vf} and on the lattice~\cite{Guo:2015tla,Abramczyk:2017oxr,Dragos:2019oxn,Alexandrou:2020mds,Bhattacharya:2021lol,Liang:2023jfj,He:2023gwp}. Current algebra estimates~\cite{Crewther:1979pi} rely on a nonvanishing CP violating pion-nucleon coupling constant $\bar{g}_{NN\pi}$. The result of this work is that $\bar{g}_{NN\pi}$ vanishes in the infinite volume. Lattice calculations do not yet provide a definitive answer, but they are consistent with our prediction (\ref{result}). The most recent work of~\cite{Liang:2023jfj}, for example, on the $24^3\times 64$ lattice, which corresponds to a probability ratio of $R_{\textrm{aver}} \approx 1$, quotes a nonvanishing value for $|d_n|$. The calculation of~\cite{Alexandrou:2020mds}, on the other hand, on the fifty times larger $64^3\times 128$ lattice volume, which corresponds to $R_{\textrm{aver}} \approx 0.13$, obtains a value consistent with zero. Both calculations show comparable errors in the raw data.

The fact that the density of zero modes decreases rapidly with increasing volume makes the zero modes undetectable by local probes in large volumes. This result should not surprise. It was stated by Leutwyler and Smilga long ago~\cite{Leutwyler:1992yt} that the phenomenon that `nontrivial topologies are accompanied by zero modes does not matter because the number of zero modes per unit volume tends to zero if $V$ tends to infinity'. One might think that this eliminates all effects of instantons and topological charge in the thermodynamic limit. This is, however, not the case. To illustrate that consider the partition function (\ref{zf}). The moments of topological charge are obtained from the free energy,
\begin{equation}
  \frac{1}{V}\, \langle Q^n\rangle_c = - i^n \left. \frac{\partial^n F(\theta)}{\partial \theta^n}\right|_{\theta=0}\,.
\end{equation}
For example, the topological susceptibility is given by
\begin{equation}
\chi_t \equiv  \frac{\langle Q^2\rangle}{V} = \left. \frac{\partial^2 F(\theta)}{\partial \theta^2}\right|_{\theta=0} = \frac{1}{2}\, \Sigma\, m 
\end{equation}
for two mass-degenerate quarks. Combining this result with the Dashen--Gell-Mann--Oakes--Renner relation at large-$N$, one arrives at the Witten-Veneziano formula for the mass of the $\eta^\prime$ meson~\cite{Witten:1979vv,Veneziano:1979ec}.
%The free energy results from the real part of the anomalous fermionic action. It is an even function of $\theta$. Thus, the measure for nontrivial topology is $\langle Q^2\rangle/V$.
Our results are not in conflict with $F(\theta)$ being an analytic function at $\theta = 0$. In summary, the absence of CP violation does not rule out any dependence on $|\theta|$, at zero and at finite temperature.

Finally, there is the question of how rigorous our statements (presently) are. To give a meaning to the $\theta$ term, a nonperturbative definition of the theory is necessary. Such a definition is given by the path integral on a compact manifold, which we have adopted. For other choices, and solutions of the problem, see~\cite{Torrieri:2020nin,Yamanaka:2022bfj,Ai:2024cnp}. The topological charge is defined by its zero modes. Unlike other definitions, it is an integer and does not have to be renormalized or stripped off its ultraviolet divergences. Exact zero modes are not a major problem for overlap fermions~\cite{Berruto:2000fx}, but they are computationally expensive, so that the problem is largely of numerical nature. The `vanishing theorem' has been overlooked for many years. Mathematically it is on the same level as the Atiyah-Singer index theorem. It is a constant in two dimensions. In four dimensions~\cite{Maier} it has been confirmed by lattice calculations~\cite{Chiu:1998bh,Ilgenfritz:2007xu,Chiu:2011dz,DiGiacomo:2015eva,Chen:2022fid}.
The `vanishing theorem' is a sufficient condition for the cluster decomposition property~\cite{Bohm:2001yx}, in the sense that it admits exactly $|Q|$ zero modes, thus forbidding mixed instanton--anti-instanton states in the semi-classical vacuum. This is a completely new development that solves a very old problem. It does not concern us directly, but should be tested on the lattice in a separate project. Everything else is based on the assumption that the topological susceptibility $\chi_t$ does not depend on the volume, and that the zero modes are ultralocal. The topological susceptibility is a measurable quantity, $\chi_t = \Sigma \, m/2$, and therefore independent of the volume. This has been confirmed by lattice calculations~\cite{DelDebbio:2004ns,Ce:2015qha}. The locality of zero modes has been analyzed in~\cite{DeGrand:2000gq,Ilgenfritz:2007xu,Ilgenfritz:2006gc,Ilgenfritz:2007ua}, albeit on small to medium sized lattices and in the quenched approximation. Our statements in Sec.\ 3 can be quantified. A favored method is the determination from the moments of the scalar density, given by the inverse participation ratio $I_p$. From $I_p$, with $p$ running from $2$ to $30$, we read a decay $|u_0(x)|^2 \, \propto \, 1/|x-x_0|^6$ from the center $x_0$~\cite{Ilgenfritz:2007xu}. This is exactly what one would expect from  an instanton backround field~\cite{Bohm:2001yx}. Instantons have a typical size of $\rho \approx 0.5\,\textrm{fm}$~\cite{Smith:1998wt}. The absence of long-range correlations can be testified by the kurtosis
\begin{equation}
  K = \frac{\langle Q^4 \rangle_c}{\langle Q^2 \rangle} = \frac{\langle Q^4 \rangle - 3 \langle Q^2\rangle^2}{\langle Q^2 \rangle} \,.
\end{equation}
If our picture is correct, we should find that $K$ gradually decreases to zero as the volume is increased. This is exactly what has been observed~\cite{Durr:2025qtq}, again in the quenched approximation. Conversely, the absence of long-range correlations can be falsified by finding a nonvanishing $K$ in the infinite volume. A Gaussian distribution is expected for a randomly distributed charge. Deviations would show up as nonzero values of the higher cumulants, $\langle Q^n \rangle_c$ with $n \geq 4$. As we have just seen, this does not seem to be the case in the large volume limit~\cite{Durr:2025qtq}. In a vacuum of sign-coherent charges, one could imagine long-range correlations, a possibility that was discussed by Leutwyler and Smilga~\cite{Leutwyler:1992yt}. They came to the conclusion that long-range correlations disappear when the volume tends to infinity, in accord with our findings. Clearly, there are holes in the lattice data, which will take time to be filled. %I am confident that the lattice calculations can be brought to a satisfactory conclusion in the intermediate future.

%A similar claim, namely that the topological charge is not observable, was made in~\cite{Yamanaka:2022bfj,Ai:2024cnp}. Even if the starting point is completely different, I broadly agree with the authors' conclusions about the implications for the topological properties of the theory. Debating the consistency of either approach is beyond the scope of this Letter.

The axion, which was postulated by Peccei and Quinn~\cite{Peccei:1977hh} to resolve the strong CP problem, has lost its justification. Even more serious, the axion extension of the Standard Model was found to be in conflict with the low energy properties of QCD~\cite{Schierholz:2023hkx}. 

The property that the vacuum divides into quantum mechanically disconnected, sign-degen\-erate sectors of topological charge $Q$ could be interpreted to mean that CP invariance is spontaneously broken. A similar phenomenon was observed in the two-dimensional nonlinear $O(3)$ model~\cite{Iwasaki:1981yq}. While this appears to have no effect on QCD, it could have implications beyond the Standard Model.

\section*{Acknowledgment}

I thank Ting-Wai Chiu for communicating his most recent results on the chirality of zero modes prior to publication. Furthermore, I am grateful to Andreas Kronfeld for reading the manuscript and for suggestions for improvement, and to Meinulf G\"ockeler for drawing my attention to Ref.~\cite{Maier}.


\begin{thebibliography}{99}

\bibitem{Abel:2020pzs}
C.~Abel, \textit{et al.}
%``Measurement of the Permanent Electric Dipole Moment of the Neutron,''
Phys. Rev. Lett. \textbf{124} (2020) 081803 [arXiv:2001.11966 [hep-ex]].  

\bibitem{Callan:1979bg}
C.~G.~Callan, R.~F.~Dashen and D.~J.~Gross,
%``Instantons as a Bridge Between Weak and Strong Coupling in \{QCD\},''
Phys. Rev. D \textbf{20} (1979) 3279.

\bibitem{tHooft:1981bkw}
G.~'t Hooft,
%``Topology of the Gauge Condition and New Confinement Phases in Nonabelian Gauge Theories,''
Nucl. Phys. B \textbf{190} (1981) 455.

\bibitem{Cardy:1981qy}
J.~L.~Cardy and E.~Rabinovici,
%``Phase Structure of Z(p) Models in the Presence of a Theta Parameter,''
Nucl. Phys. B \textbf{205} (1982) 1.

\bibitem{Nakamura:2019ind}
Y.~Nakamura and G.~Schierholz,
%``Does confinement imply CP invariance of the strong interactions?,''
PoS \textbf{LATTICE2019} (2019) 172
[arXiv:1912.03941 [hep-lat]].

\bibitem{Nakamura:2021meh}
Y.~Nakamura and G.~Schierholz,
%``The strong CP problem solved by itself due to long-distance vacuum effects,''
Nucl. Phys. B \textbf{986} (2023) 116063
[arXiv:2106.11369 [hep-ph]].

\bibitem{Schierholz:2025tns}
G.~Schierholz,
%``Absence of CP Violation in the Strong Interaction: Vacuum thwarts Axion,''
PoS \textbf{LATTICE2024} (2025) 398
[arXiv:2502.04092 [hep-lat]].

\bibitem{Vafa:1984xg}
C.~Vafa and E.~Witten,
%``Parity Conservation in QCD,''
Phys. Rev. Lett. \textbf{53} (1984) 535.

\bibitem{Schierholz:2022wuc}
G.~Schierholz,
%``Dynamical solution of the strong CP problem within QCD?,''
EPJ Web Conf. \textbf{274} (2022) 01009
[arXiv:2212.05485 [hep-lat]].

\bibitem{Matsui:1986dk}
T.~Matsui and H.~Satz,
%``$J/\psi$ Suppression by Quark-Gluon Plasma Formation,''
Phys. Lett. B \textbf{178} (1986) 416.

\bibitem{Fujikawa:1980eg}
K.~Fujikawa,
%``Path Integral for Gauge Theories with Fermions,''
Phys. Rev. D \textbf{21} (1980) 2848
[erratum: Phys. Rev. D \textbf{22} (1980) 1499].

\bibitem{Hasenfratz:1998ri}
P.~Hasenfratz, V.~Laliena and F.~Niedermayer,
%``The Index theorem in QCD with a finite cutoff,''
Phys. Lett. B \textbf{427} (1998) 125
[arXiv:hep-lat/9801021 [hep-lat]].

\bibitem{Neuberger:1997fp}
H.~Neuberger,
%``Exactly massless quarks on the lattice,''
Phys. Lett. B \textbf{417} (1998) 141
[arXiv:hep-lat/9707022 [hep-lat]];
%``More about exactly massless quarks on the lattice,''
Phys. Lett. B \textbf{427} (1998) 353
[arXiv:hep-lat/9801031 [hep-lat]].

\bibitem{DeGrand:2000gq}
T.~A.~DeGrand and A.~Hasenfratz,
%``Low lying fermion modes, topology and light hadrons in quenched QCD,''
Phys. Rev. D \textbf{64} (2001) 034512
[arXiv:hep-lat/0012021 [hep-lat]].

\bibitem{Ansourian:1977qe}
M.~M.~Ansourian,
%``Index Theory and the Axial Current Anomaly in Two-Dimensions,''
Phys. Lett. B \textbf{70} (1977) 301.

\bibitem{Nielsen:1977aw}
N.~K.~Nielsen and B.~Schroer,
%``Axial Anomaly and Atiyah-Singer Theorem,''
Nucl. Phys. B \textbf{127} (1977) 493.

\bibitem{Brown:1977bj}
L.~S.~Brown, R.~D.~Carlitz and C.~k.~Lee,
%``Massless Excitations in Instanton Fields,''
Phys. Rev. D \textbf{16} (1977) 417.

\bibitem{Carlitz:1978xu}
R.~D.~Carlitz and C.~k.~Lee,
%``Physical Processes in Pseudoparticle Fields: The Role of Fermionic Zero Modes,''
Phys. Rev. D \textbf{17} (1978) 3238.

\bibitem{Maier}
S.~Maier,
%``Generic Metrics and Connections on Spin- and Spin^c-Manifolds,''
Commun. Math. Phys. \textbf{188} (1997) 407.

\bibitem{Leutwyler:1992yt}
H.~Leutwyler and A.~V.~Smilga,
%``Spectrum of Dirac operator and role of winding number in QCD,''
Phys. Rev. D \textbf{46} (1992) 5607.

\bibitem{Chiu:1998bh}
T.~W.~Chiu,
%``The Spectrum and topological charge of exactly massless fermions on the lattice,''
Phys. Rev. D \textbf{58} (1998) 074511
[arXiv:hep-lat/9804016 [hep-lat]].

\bibitem{Ilgenfritz:2007xu}
E.~M.~Ilgenfritz, K.~Koller, Y.~Koma, G.~Schierholz, T.~Streuer and V.~Weinberg,
%``Exploring the structure of the quenched QCD vacuum with overlap fermions,''
Phys. Rev. D \textbf{76} (2007) 034506
[arXiv:0705.0018 [hep-lat]].

\bibitem{Chiu:2011dz}
T.~W.~Chiu, T.~H.~Hsieh and Y.~Y.~Mao,
%``Topological Susceptibility in Two Flavors Lattice QCD with the Optimal Domain-Wall Fermion,''
Phys. Lett. B \textbf{702} (2011) 131
[arXiv:1105.4414 [hep-lat]]; T.~W.~Chiu,
%``First study of $N_f=2+1+1$ lattice QCD with physical domain-wall quarks,''
PoS \textbf{LATTICE2019} (2020) 133
[arXiv:2002.06126 [hep-lat]].

\bibitem{DiGiacomo:2015eva}
A.~Di Giacomo and M.~Hasegawa,
%``Instantons and Monopoles,''
Phys. Rev. D \textbf{91} (2015) 054512
[arXiv:1501.06517 [hep-lat]].

\bibitem{Chen:2022fid}
Y.~C.~Chen, T.~W.~Chiu and T.~H.~Hsieh,
%``Topological susceptibility in finite temperature QCD with physical (u/d,s,c) domain-wall quarks,''
Phys. Rev. D \textbf{106} (2022) 074501
[arXiv:2204.01556 [hep-lat]].

\bibitem{Chiu}
T.~W.~Chiu, private communication.

\bibitem{Luscher:2010iy}
M.~L\"uscher,
%``Properties and uses of the Wilson flow in lattice QCD,''
JHEP \textbf{08} (2010) 071
[erratum: JHEP \textbf{03} (2014) 092]
[arXiv:1006.4518 [hep-lat]].

\bibitem{Luscher:1979yd}
M.~L\"uscher,
%``EXACT INSTANTON GASES,''
NATO Sci. Ser. B \textbf{59} (1980) 205-216.

\bibitem{Bohm:2001yx}
M.~B\"ohm, A.~Denner and H.~Joos,
{\it Gauge theories of the strong and electroweak interaction} (Teubner, Stuttgart, 2001).

\bibitem{Iwasaki:1981yq}
Y.~Iwasaki,
%``Instanton Contributions and the Cluster Property of the Vacuum,''
Phys. Lett. B \textbf{104} (1981) 458.

\bibitem{Guadagnoli:2002nm}
D.~Guadagnoli, V.~Lubicz, G.~Martinelli and S.~Simula,
%``Neutron electric dipole moment on the lattice: A Theoretical reappraisal,''
JHEP \textbf{04} (2003) 019
[arXiv:hep-lat/0210044 [hep-lat]].

\bibitem{Koma:2005sw}
Y.~Koma, E.~M.~Ilgenfritz, K.~Koller, G.~Schierholz, T.~Streuer and V.~Weinberg,
%``Localization properties of the topological charge density and the low lying eigenmodes of overlap fermions,''
PoS \textbf{LATTICE2005} (2006) 300
[arXiv:hep-lat/0509164 [hep-lat]].

\bibitem{Ilgenfritz:2006gc}
E.~M.~Ilgenfritz, K.~Koller, Y.~Koma, G.~Schierholz, T.~Streuer and V.~Weinberg,
%``Vacuum structure as seen by overlap fermions,''
AIP Conf. Proc. \textbf{892} (2007) 187
[arXiv:hep-lat/0611007 [hep-lat]].

\bibitem{Ilgenfritz:2007ua}
E.~M.~Ilgenfritz, K.~Koller, Y.~Koma, G.~Schierholz, T.~Streuer, V.~Weinberg and M.~Quandt,
%``Localization of overlap modes and topological charge, vortices and monopoles in SU(3) LGT,''
PoS \textbf{LATTICE2007} (2007) 311
[arXiv:0710.2607 [hep-lat]].

\bibitem{Ilgenfritz:2009qw}
E.~M.~Ilgenfritz, K.~Koller, Y.~Koma, G.~Schierholz and V.~Weinberg,
%``Topological Structure of the QCD Vacuum Revealed by Overlap Fermions,''
arXiv:0912.2281 [hep-lat].

\bibitem{Alexandrou:2020mds}
C.~Alexandrou, A.~Athenodorou, K.~Hadjiyiannakou and A.~Todaro,
%``Neutron electric dipole moment using lattice QCD simulations at the physical point,''
Phys. Rev. D \textbf{103} (2021) 054501
[arXiv:2011.01084 [hep-lat]].

\bibitem{Crewther:1979pi}
R.~J.~Crewther, P.~Di Vecchia, G.~Veneziano and E.~Witten,
%``Chiral Estimate of the Electric Dipole Moment of the Neutron in Quantum Chromodynamics,''
Phys. Lett. B \textbf{88} (1979) 123
[erratum: Phys. Lett. B \textbf{91} (1980) 487].

\bibitem{Shifman:1979if}
M.~A.~Shifman, A.~I.~Vainshtein and V.~I.~Zakharov,
%``Can Confinement Ensure Natural CP Invariance of Strong Interactions?,''
Nucl. Phys. B \textbf{166} (1980) 493.

\bibitem{mathematica} Wolfram Research, Inc., Mathematica, Version 11.0 (2016, Champaign, USA).

\bibitem{Guo:2012vf}
F.~K.~Guo and U.~G.~Meissner,
%``Baryon electric dipole moments from strong CP violation,''
JHEP \textbf{12} (2012) 097
[arXiv:1210.5887 [hep-ph]].

\bibitem{Guo:2015tla}
F.~K.~Guo, R.~Horsley, U.~G.~Meissner, Y.~Nakamura, H.~Perlt, P.~E.~L.~Rakow, G.~Schierholz, A.~Schiller and J.~M.~Zanotti,
%``The electric dipole moment of the neutron from 2+1 flavor lattice QCD,''
Phys. Rev. Lett. \textbf{115} (2015) 062001
[arXiv:1502.02295 [hep-lat]].

\bibitem{Abramczyk:2017oxr}
M.~Abramczyk, S.~Aoki, T.~Blum, T.~Izubuchi, H.~Ohki and S.~Syritsyn,
%``Lattice calculation of electric dipole moments and form factors of the nucleon,''
Phys. Rev. D \textbf{96} (2017) 014501
[arXiv:1701.07792 [hep-lat]].

\bibitem{Dragos:2019oxn}
J.~Dragos, T.~Luu, A.~Shindler, J.~de Vries and A.~Yousif,
%``Confirming the Existence of the strong CP Problem in Lattice QCD with the Gradient Flow,''
Phys. Rev. C \textbf{103} (2021) 015202
[arXiv:1902.03254 [hep-lat]].

\bibitem{Bhattacharya:2021lol}
T.~Bhattacharya, V.~Cirigliano, R.~Gupta, E.~Mereghetti and B.~Yoon,
%``Contribution of the QCD $\Theta$-term to the nucleon electric dipole moment,''
Phys. Rev. D \textbf{103} (2021) 114507
[arXiv:2101.07230 [hep-lat]].

\bibitem{Liang:2023jfj}
J.~Liang, A.~Alexandru, T.~Draper, K.~F.~Liu, B.~Wang, G.~Wang and Y.~B.~Yang, 
%``Nucleon electric dipole moment from the \ensuremath{\theta} term with lattice chiral fermions,''
Phys. Rev. D \textbf{108} (2023) 094512
[arXiv:2301.04331 [hep-lat]].

\bibitem{He:2023gwp}
F.~He, M.~Abramczyk, T.~Blum, T.~Izubuchi, H.~Ohki and S.~Syritsyn,
%``The calculations of Nucleon Electric Dipole Moment using background field on Lattice QCD,''
[arXiv:2311.06106 [hep-lat]].

\bibitem{Witten:1979vv}
E.~Witten,
%``Current Algebra Theorems for the U(1) Goldstone Boson,''
Nucl. Phys. B \textbf{156} (1979) 269.

\bibitem{Veneziano:1979ec}
G.~Veneziano,
%``U(1) Without Instantons,''
Nucl. Phys. B \textbf{159} (1979) 213.

\bibitem{Torrieri:2020nin}
G.~Torrieri and H.~D.~Truran,
%``The strong CP problem, general covariance, and horizons,''
Class. Quant. Grav. \textbf{38} (2021) 215002
[arXiv:2007.13183 [hep-th]].

\bibitem{Yamanaka:2022bfj}
N.~Yamanaka,
%``Unobservability of the topological charge in nonabelian gauge theory: Ward-Takahashi identity and phenomenological aspects,''
[arXiv:2212.11820 [hep-ph]].

\bibitem{Ai:2024cnp}
W.~Y.~Ai, B.~Garbrecht and C.~Tamarit,
%``CP Conservation in the Strong Interactions,''
Universe \textbf{10} (2024) 189
[arXiv:2404.16026 [hep-ph]].

\bibitem{Berruto:2000fx}
F.~Berruto, R.~Narayanan and H.~Neuberger,
%``Exact local fermionic zero modes,''
Phys. Lett. B \textbf{489} (2000) 243
[arXiv:hep-lat/0006030 [hep-lat]].

\bibitem{DelDebbio:2004ns}
L.~Del Debbio, L.~Giusti and C.~Pica,
%``Topological susceptibility in the SU(3) gauge theory,''
Phys. Rev. Lett. \textbf{94} (2005) 032003
[arXiv:hep-th/0407052 [hep-th]].

\bibitem{Ce:2015qha}
M.~C\`e, C.~Consonni, G.~P.~Engel and L.~Giusti,
%``Non-Gaussianities in the topological charge distribution of the SU(3) Yang--Mills theory,''
Phys. Rev. D \textbf{92} (2015) 074502
[arXiv:1506.06052 [hep-lat]].

\bibitem{Smith:1998wt}
D.~A.~Smith and M.~J.~Teper,
%``Topological structure of the SU(3) vacuum,''
Phys. Rev. D \textbf{58} (1998) 014505
[arXiv:hep-lat/9801008 [hep-lat]].

%\bibitem{Hernandez:1998et}
%P.~Hernandez, K.~Jansen and M.~L\"uscher,
%%``Locality properties of Neuberger's lattice Dirac operator,''
%Nucl. Phys. B \textbf{552} (1999) 363
%[arXiv:hep-lat/9808010 [hep-lat]].

\bibitem{Durr:2025qtq}
S.~D\"urr and G.~Fuwa,
%``The topological susceptibility and excess kurtosis in SU(3) Yang-Mills theory,''
arXiv:2501.08217 [hep-lat].

\bibitem{Peccei:1977hh}
R.~D.~Peccei and H.~R.~Quinn,
%``CP Conservation in the Presence of Instantons,''
Phys. Rev. Lett. \textbf{38} (1977) 1440.

\bibitem{Schierholz:2023hkx}
G.~Schierholz,
%``Repercussions of the Peccei-Quinn axion on QCD,''
doi:10.1142/S021773232430009X
[arXiv:2307.08310 [hep-ph]].

\end{thebibliography}
\end{document}